# Experimental Test of Universal Conductance Fluctuations by means of Wave-Chaotic Microwave Cavities.

Sameer Hemmady[1,2,3,4], James Hart[1], Xing Zheng[1], Thomas M. Antonsen Jr.[1,2,3], Edward Ott[1,2,3] and Steven M. Anlage[1,2,4].
[1]Department of Physics, University of Maryland, College Park, MD 20742-4111, U.S.A.
[Dated : September 13th, 2006].

## Abstract:

The mathematical equivalence of the time-independent Schrödinger equation and the Helmholtz equation is exploited to provide a novel means of studying universal conductance fluctuations in ballistic chaotic mesoscopic systems using a two-dimensional microwave-cavity. The classically chaotic ray trajectories within a suitably-shaped microwave cavity play a role analogous to that of the chaotic dynamics of non-interacting electron transport through a ballistic quantum dot in the absence of thermal fluctuations. The microwave cavity is coupled through two single-mode ports and the effect of non-ideal coupling between the ports and cavity is removed by a previously developed method based on the measured radiation impedance matrix. The Landauer-Büttiker formalism is applied to obtain the conductance of a corresponding mesoscopic quantum-dot device. We find good agreement for the probability density functions (PDFs) of the experimentally derived surrogate conductance, as well as its mean and variance, with the theoretical predictions of Brouwer and Beenakker. We also observe a linear relation between the quantum dephasing parameter and the cavity ohmic loss parameter.

## I. Introduction

Much attention has been focused on the problem of mesoscopic transport through a quantum dot in which a two-dimensional electron gas system contained within an arbitrarily-shaped potential-well boundary is connected to two electron reservoirs through leads– the source ($s$) and drain ($d$). Recently it has been possible to fabricate quantum dots with low impurity content where the elastic mean free paths of the enclosed electrons are typically much larger than the physical size of the dot [1]. Electron transport through such "ballistic dots" is governed by elastic collisions off the enclosing potential-well boundaries. It has been observed that the terminal conductance of such dots, defined as $\hat{G} = I_s/(V_s - V_d)$ where $I_s$ is the source current flowing into the dot and $(V_s - V_d)$ is the potential difference between these two leads, exhibits strong, reproducible fluctuations on the order of the quantum of conductance ($G_0 = e^2/h$) [2, 3, 4]. These fluctuations arise from quantum-interference effects due to the phase-coherent electron transport within such dots and have been explained using the hypothesis that the fluctuations are governed by Random Matrix Theory [5]. Similar universal conductance fluctuations (UCF) were previously observed in other systems such as quasi-one-dimensional metal wires [6, 7, 8].

In a quantum dot, this phase coherence is partly lost by opening the system to the outside world during the process of measurement of the conductance. Quantum phase decoherence (dephasing) can also be induced due to the presence of impurities within the dot, thermal fluctuations, or electron-electron interactions, all of which lead to more classical properties for electron transport [9]. Significant theoretical and experimental effort has been devoted to studying the dephasing of the transport electrons in quantum dots [10, 11, 12]. One class of theoretical dephasing models utilizes a "fictitious voltage probe ($\phi$)" attached to the dot that has a number of channels (modes) $N_\phi$, each of which contains a tunnel-barrier with transmission probability $\Gamma_\phi$ for the electrons that enter the channel from the dot. Electrons that enter one of the modes of this probe are re-injected into the dot with a phase that is uncorrelated with their initial phase, and there is no net current through the fictitious probe. An alternative model of electron transport employs a uniform imaginary term in the electron potential [13, 14], leading to loss of probability density with time. It was shown that [15], as far as the conductance is concerned, these two models yield equivalent predictions in the limit when the number of channels in the dephasing lead $N_\phi \to \infty$ and $\Gamma_\phi \to 0$, with the product $\gamma = N_\phi \Gamma_\phi$ remaining finite ("the locally weak absorbing limit") [16]. A similar idea exists for describing ohmic losses in the microwave cavity in terms of non-ideally coupled "parasitic channels" **[17]**. Since the ohmic losses in the microwave cavity are to good approximation uniformly distributed, we can make use of the equivalence of the imaginary potential and voltage leads models mentioned above to relate the de-phasing parameter employed by electron-transport theory **[15]** to the loss parameter of our microwave cavity ($k^2/(\Delta k_n^2 Q)$) **[18, 19, 20, 21]**. Here, $k = 2\pi f/c$ is the wavenumber for the incoming frequency $f$ and $\Delta k_n^2$ is the mean-spacing of the adjacent eigenvalues of the Helmholtz operator, $\nabla^2 + k^2$, as predicted by the Weyl Formula **[22]** for the closed system. The quantity Q represents the loaded quality factor of the cavity. Using the prescription outlined by Ref. **[15]** we can directly determine the analog of conductance for the microwave cavity and make detailed comparisons of data to theory.

[2] Dept. of Electrical and Computer Engineering.
[3] Institute for Research in Electronics and Applied Physics.
[4] Center for Superconductivity Research.

We use an electromagnetic analog of a quantum dot in the form of a two-dimensional chaotic microwave resonator **[23]**. In the case of a cavity thin in one dimension, Maxwell's equations reduce to a two-dimensional scalar Helmholtz equation. Owing to the analogy between the scalar Helmholtz equation and the Schrödinger equation **[24]**, the chaotic microwave cavity is an ideal surrogate for a ballistic quantum dot without the complicating effects of thermal fluctuations **[4]**, Coulomb interactions, or impurities. The microwave cavity is driven by two ports, both of which support a single propagating mode and are analogous to the source and drain leads in the quantum dot. The microwave analog permits detailed measurements of the eigenvalues **[25, 26, 27]**, eigenfunctions **[28, 29, 30, 31, 32]**, scattering and reaction matrices **[19, 20, 21, 33, 34, 35, 36]**, in a system where every detail of the potential and the coupling channels can be controlled.

Adopting a variant of the Landaeur-Buttiker formalism, the normalized conductance ($G = \hat{G}/2G_o$) can be expressed in terms of the scattering matrix $\vec{s} = \begin{pmatrix} s_{11} & s_{12} \\ s_{21} & s_{22} \end{pmatrix}$ of a chaotic cavity when the leads (ports) are perfectly coupled to the cavity **[37]**. i.e.,

$$G = |s_{12}|^2 + \frac{(1-|s_{11}|^2-|s_{12}|^2)(1-|s_{22}|^2-|s_{21}|^2)}{2-|s_{11}|^2-|s_{12}|^2-|s_{21}|^2-|s_{22}|^2}, \quad (1)$$

where the first term describes the direct (phase coherent) transport through the microwave-cavity and corresponds to the conductance of the quantum-dot due to the electrons that did not enter the fictitious voltage probe. The second term is a correction that describes the conductance due to the electrons that are re-injected into the dot from the phase-breaking fictitious voltage-probe, thereby ensuring particle conservation in the voltage-probe model **[15]**.

In the time-reversal symmetric case with single-mode leads, Ref. **[15]** has shown that as $\gamma$ increases the probability density function of $G$ (i.e. $P(G;\gamma)$) becomes more and more sharply peaked around the classical value of $G = 1/2$. In the limit of large $\gamma$, an asymptotic analytic expression for $P(G;\gamma)$ is predicted to be **[15]**,

$$P(G;\gamma) = \frac{1}{2}\gamma(1+|x|-x)e^{-|x|}$$
with $x = 2\gamma(G-1/2)$. (2)

This yields a large-$\gamma$ asymptotic expression for the mean and variance of $G$ which are predicted to be **[15]**,

$$<G> = \frac{1}{2} - \frac{1}{2\gamma} + O(\gamma^{-2}), \quad (3)$$

$$\text{var}(G) = \frac{3}{4\gamma^2} + O(\gamma^{-3}). \quad (4)$$

This paper is divided into the following sections. In section II, we present a brief description of the experimental setup and data analysis. In section III, we present the experimental results by first examining the relation between the dephasing parameter ($\gamma$) and the cavity loss parameter ($k^2/(\Delta k_n^2 Q)$). Section IV then uncovers the PDFs of the experimentally determined universal conductance fluctuations for increasing values of $\gamma$ and compares them with predictions from Ref.**[15]** and Random Matrix Theory. In Section V, we experimentally test the predictions from Ref.**[15]** for the mean and variance of these universal conductance fluctuation PDFs as a function of $\gamma$. Finally, section VI concludes this paper with a summary of our experimental findings and its implications.

## II. Experimental Setup and Data Analysis

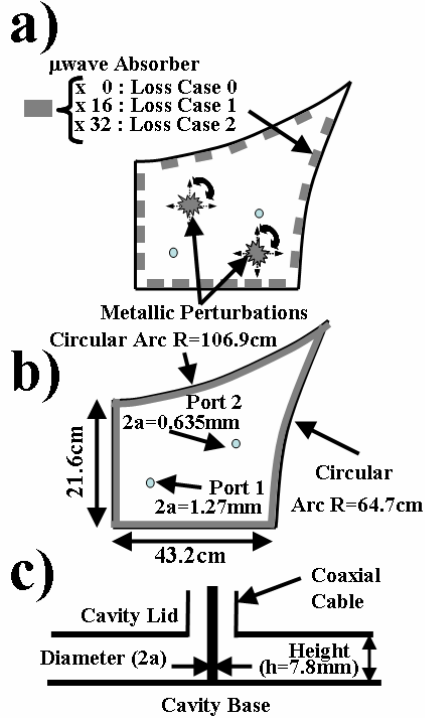

Fig.1: (a) Top view of quarter-bow-tie microwave cavity used for the experimental "Cavity Case". The two perturbations are shown as the gray shapes. The small, gray, uniformly-spaced rectangles lining the side walls of the cavity represent 2cm-long strips of microwave absorber which are used to control the loss in the cavity.(Loss Case 0 : 0 strips, Loss Case 1: 16 strips, Loss Case 2: 32 strips). (b) The implementation of the experimental "Radiation Case" is shown. The gray lining on the side walls is a homogenous layer of microwave absorber about 2 mm thick. The physical dimensions of the cavity are shown in the schematic. The approximate locations of the two driving-ports are also shown. (c) Cross-section view of both driving-ports inside the cavity. The cavity is 7.87 mm in depth. The diameter of the inner conductor is 2a (=1.27 mm for Port 1; =0.635 mm for Port 2).

The microwave cavity under study is a metallic, air-filled, quarter bow-tie shaped chaotic resonator (Fig. 1(a)) which is quasi-two-dimensional for frequencies below 19.05 GHz. The cavity is driven by two single-mode, coaxial transmission lines whose inner conductor (diameters 2a=1.27mm for Port 1, 2a=0.635mm for Port 2 as shown in Fig. 1(b)) extends from the top plate of the cavity and makes contact with the bottom plate (shown as schematic in Fig.1(c)). An ensemble data set of one-hundred similar cavities with different internal field configurations is generated by rotating and translating two metallic perturbations, each of which are roughly the size of a wavelength at 5 GHz (gray jagged shapes in Fig. 1(a)), within the cavity volume. This approach of configuration averaging to approximate a pure ensemble average is similar in principle to deforming the shape of the potential-well boundary of a quantum dot as performed by [4] although in our case the volume of the system is fixed. In addition to the intrinsic ohmic loss in the cavity, the degree of loss can be further increased in a controlled manner by partially lining the inner side-walls of the cavity with 2 cm-long strips of microwave absorber having uniform spacing. This results in three experimental Loss Cases - Loss Case 0: no absorbing strips, Loss Case 1: 16 absorbing strips, Loss Case 2: 32 absorbing strips. A fourth experimental Loss Case is created by placing the Loss Case 0 cavity in a bath of dry-ice (solid $CO_2$ at $-78.5°C$). This has the effect of slightly increasing the Loss Case 0 cavity Q value (by ~10%). We refer to this case as the "dry-ice case". A more detailed explanation of our experimental setup and data analysis can be found in [21].

For our investigation, we generate a large ensemble of 2x2 cavity scattering matrices ($\vec{S}$) for the four loss cases through measurements for different configurations of the perturbers at many frequencies in a range from 3 to 18 GHz (covering about 800 modes of the cavity). Using the "radiation impedance" approach [18, 19, 20, 21] (Fig.1(b)), the non-ideal coupling details of the two ports are removed to yield an ensemble of normalized 2x2 scattering matrices ($\vec{s}$) from which the conductance statistics are derived using Eq. (1).

Prior to reporting results for different "data sets" where each data set corresponds to one of our four loss cases and a frequency range typically spanning about 1 GHz, we estimate the value of $\gamma$ for each data set. We derive an analytic expression for the mean value of the absorption probability $\langle T \rangle$ in terms of $\gamma$ from Eq. (17a) of [15],

$$\langle T_1 \rangle = \langle T_2 \rangle = \langle T \rangle = \frac{1}{4\gamma}(e^{-\gamma}(4(e^\gamma - \gamma - 1) \\ + 4e^\gamma(2e^\gamma - 2 - \gamma(2+\gamma))\xi(-\gamma) \quad (5) \\ - 2e^{\gamma/2}(e^\gamma(2+\gamma(\gamma-2))-2)\xi(-\gamma/2)))$$

where $\xi(z) = -\int_{-z}^{\infty}\frac{e^{-t}}{t}dt$ is the exponential integral function. Here $1-T_1$ and $1-T_2$ are the eigenvalues of $\vec{\vec{s}}\vec{\vec{s}}^{\dagger}$, and $\langle T_1 \rangle = \langle T_2 \rangle$ since the joint PDF of $T_1$ and $T_2$ (Eq. (17a) of **[15]**) is symmetrically distributed. The inset in Fig.2 shows the relation between $\langle T \rangle$ and $\gamma$ on a semi-log plot (blue curve). By determining the value of $\langle T \rangle$ from the measured data set, Eq.(5) then uniquely determines the corresponding value of $\gamma$ ($\equiv \gamma_{\langle T \rangle}$).

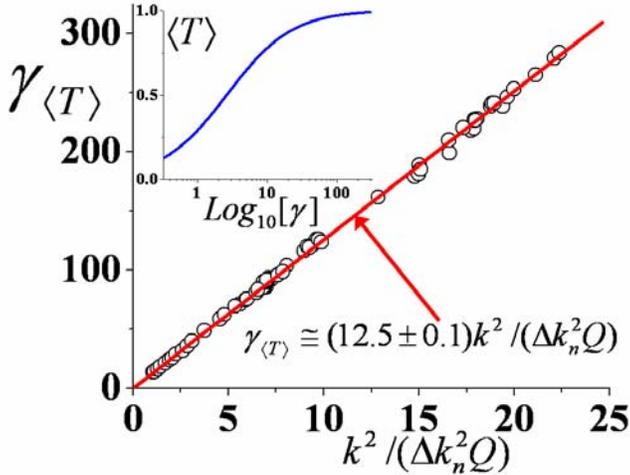

Fig.2: The relation between the experimentally determined value for $\gamma_{\langle T \rangle}$ and $k^2/(\Delta k_n^2 Q)$ corresponding to each 1GHz wide data-set over the frequency range 3 GHz to 18 GHz for the different cavity Loss Cases-0, 1 and 2, is shown as the black circles. A linear fit (red line) yields the empirical expression $\gamma_{\langle T \rangle} = (12.5 \pm 0.1) k^2/(\Delta k_n^2 Q)$. **Inset:** The relation between $\langle T \rangle$ and $\gamma$ as expressed in Eq. (5) is shown as the blue curve on a semi-log scale.

To determine the cavity loss parameter $k^2/(\Delta k_n^2 Q)$ for our data sets, we employ one of two procedures. For data sets with $k^2/(\Delta k_n^2 Q) \leq 5$, we numerically generate marginal PDFs of the real and imaginary parts of the normalized impedance ($\vec{\vec{z}} = (\vec{\vec{s}} + \vec{\vec{1}})(\vec{\vec{s}} - \vec{\vec{1}})^{-1}$) eigenvalues using random-matrix Monte-Carlo simulations with square matrices of size $N = 1000$, and the value of $k^2/(\Delta k_n^2 Q)$ in the simulations ranges from 0.1 to 5 in steps of 0.1. We determine the variance ($\sigma^2$) of these numerically generated PDFs and fit it to a polynomial function $\sigma^2 = \Theta(k^2/(\Delta k_n^2 Q))$ of high order. We then determine the variances of the real and imaginary parts of each experimental data set and solve the inverse polynomial function $k^2/(\Delta k_n^2 Q) = \Theta^{-1}(\sigma^2)$ to obtain a unique estimate of $k^2/(\Delta k_n^2 Q)$ corresponding to each data set. For data sets with $k^2/(\Delta k_n^2 Q) > 5$, we use the relation $\sigma^2 = \dfrac{1}{\pi(k^2/(\Delta k_n^2 Q))}$ **[18, 21, 38]** which has been validated experimentally in Ref. **[19]**.

### III. Relationship between the dephasing parameter ($\gamma$) and the cavity loss parameter ($k^2/(\Delta k_n^2 Q)$)

We begin by examining the relationship between the estimated dephasing parameter $\gamma_{\langle T \rangle}$ and the estimated cavity loss parameter $k^2/(\Delta k_n^2 Q)$. By employing a sliding frequency window 1 GHz wide that runs over each of the three Loss Cases - 0, 1, 2, from 3 GHz to 18 GHz, we estimate the value of $\gamma_{\langle T \rangle}$ and the corresponding value of $k^2/(\Delta k_n^2 Q)$ for each window. The comparison is shown as the black circles in Fig.2. A linear fit (red line in Fig. 2) yields the empirical expression $\gamma_{\langle T \rangle} = (12.5 \pm 0.1) k^2/(\Delta k_n^2 Q)$ for about 70 points with values for $\gamma_{\langle T \rangle}$ ranging from about 11 to about 300. By comparing the Poynting theorem for the electromagnetic cavity with the continuity equation for the probability density in the quantum system **[39]**, we find $\gamma = 4\pi k^2/(\Delta k_n^2 Q)$, with $4\pi = 12.56....$ This result can be considered an empirical confirmation of the proposed equivalence of the imaginary potential (uniform volume losses) and de-phasing lead models in the limit considered in **[15]**. The 1 GHz width of our sliding window was chosen to be large enough to overcome the effects of short-ray paths (which are not removed by only configuration averaging **[18, 21]**), but at the same time small enough that the cavity losses can be assumed to be approximately constant over this frequency range.

### IV. Uncovering the Universal Conductance Fluctuation PDFs

In Fig. 3, the experimentally obtained histogram approximation (symbols) to the probability density functions (PDFs) of the normalized conductance ($P(G;\gamma)$) derived from the normalized scattering matrix $\vec{\vec{s}}$ and Eq. (1) is shown for four cavity data sets. The colored solid lines (magenta, black, green, red) are the asymptotic analytic expression for $P(G,\gamma)$ (Eq. (2)) with values of $\gamma$ that correspond to the estimated $\gamma_{\langle T \rangle}$ values obtained from the four cavity data sets. The blue-colored solid line in Fig. 3(a) is a random matrix Monte-Carlo simulation for values of $\gamma_{\langle T \rangle}$ corresponding to the data set in Fig. 3(a). The red error bars (roughly the size of the symbols) in Fig. 3 which are representative of the typical statistical binning error of the experimental histograms show that the agreement between the data (shown by the symbols) and the theoretical predictions (shown by the solid curves) improves as the value of $\gamma_{\langle T \rangle}$ increases. This is to be expected as Eq. (2) is valid only in the high dephasing limit ($\gamma \gg 1$). Similar good agreement between the data and Eq. (2) is obtained for all of the ~ 40 data sets that we examined in which the frequency ranges and cavity loss cases resulted in the $\gamma_{\langle T \rangle}$ parameter greater than ~ 18.

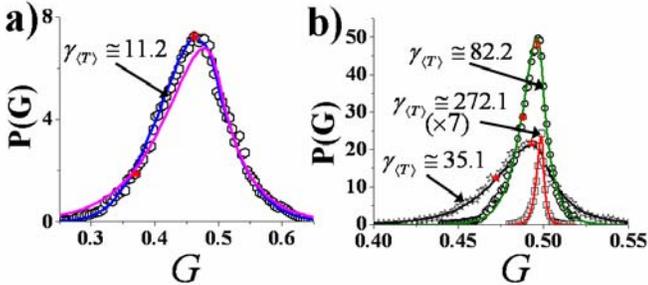

Fig.3: (a) PDFs for the normalized conductance $P(G;\gamma)$ obtained from a chaotic cavity for the dry-ice case : 4.1-4.7 GHz (hexagons) ($k^2/(\Delta k_n^2 Q) = 0.9 \pm 0.1 ; \gamma_{\langle T \rangle} = 11.2 \pm 0.1$) and in (b) Loss Case 0 : 16.8-17.6 GHz (stars) ($k^2/(\Delta k_n^2 Q) = 2.8 \pm 0.1 ; \gamma_{\langle T \rangle} = 35.1 \pm 0.1$); Loss Case 1 : 8.3-9.5 GHz (circles) ($k^2/(\Delta k_n^2 Q) = 6.6 \pm 0.1 ; \gamma_{\langle T \rangle} = 82.2 \pm 0.1$) and Loss Case 2 : 16.8-17.6 GHz (squares) ($k^2/(\Delta k_n^2 Q) = 21.7 \pm 0.1 ; \gamma_{\langle T \rangle} = 272.1 \pm 0.1$). The red error bars (which are roughly the size of the symbols) are indicative of the typical statistical binning error in the experimentally determined normalized conductance PDFs. The magenta, black, green and red lines are obtained from Eq. (2) and correspond to $\gamma$-values of 11.2, 35.1, 82.2 and 272.1 respectively. The blue line in (a) is obtained from Random Matrix Monte Carlo simulation corresponding to a $\gamma$-value of 11.2.

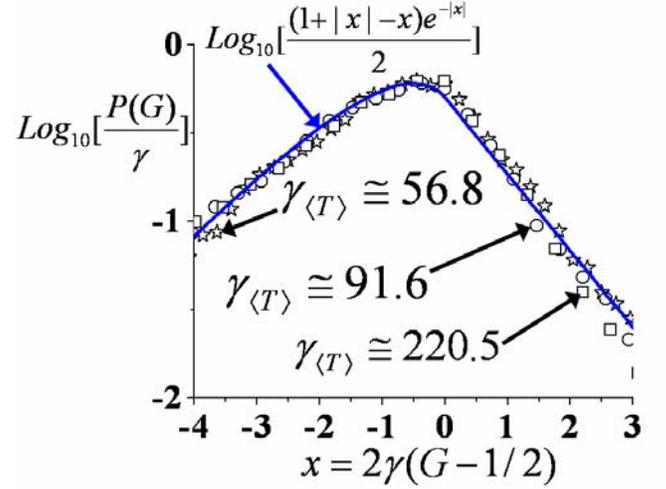

Fig.4: The universal scaling behavior of the conductance distributions is shown. The vertical-axis represents $Log_{10}[\frac{P(G,\gamma)}{\gamma}]$ with the corresponding $x = 2\gamma(G-1/2)$ along the horizontal-axis for three representative data sets consisting of Loss Case 1 : 5.01-6.08 GHz (stars) ($k^2/(\Delta k_n^2 Q) = 4.5 \pm 0.1 ; \gamma_{\langle T \rangle} = 56.6 \pm 0.1$); Loss Case 1 : 13.6-14.6 GHz (circles) ($k^2/(\Delta k_n^2 Q) = 7.3 \pm 0.1 ; \gamma_{\langle T \rangle} = 91.6 \pm 0.1$) and Loss Case 2 : 13.6-14.6 GHz (squares) ($k^2/(\Delta k_n^2 Q) = 17.7 \pm 0.1 ; \gamma_{\langle T \rangle} = 220.5 \pm 0.1$). The blue line is Eq.(2).

In order to bring out the universal scaling behavior of the $P(G;\gamma)$ distributions (Eq.(2)) and also to test that these distributions remain strictly non-Gaussian for increasing values of $\gamma$ (as predicted by [15]), we rescale the $P(G;\gamma)$ distributions by plotting $Log_{10}[\frac{P(G;\gamma)}{\gamma}]$ versus $x = 2\gamma(G-1/2)$ in Fig. 4 for three representative data sets with $\gamma_{\langle T \rangle}$ ranging from about 56 to about 220. We observe that the three data sets roughly fall on top of each other. The solid blue curve is the theoretical prediction, Eq.(2) which is in good agreement with the data. We observe some deviation of the symbols from the theoretical curve near $x \cong +2$. This is attributed to the lack of adequate statistics in the tails of the experimentally determined histogram approximations to the probability density functions of the conductance. Overall, for values of $x$ ranging from $-4$ to $+2$, the agreement is qualitatively good and applies over other data sets where $\gamma_{\langle T \rangle}$ ranges from about 18 to about 330. The asymmetric (non-parabolic) nature of the experimental data, represented by the symbols, confirms that the experimentally obtained $P(G;\gamma)$ remains strictly non-Gaussian and negatively skewed even for large values of $\gamma$ as predicted by [15].

## V. Testing Predictions for the Mean and Variance of the UCF PDFs

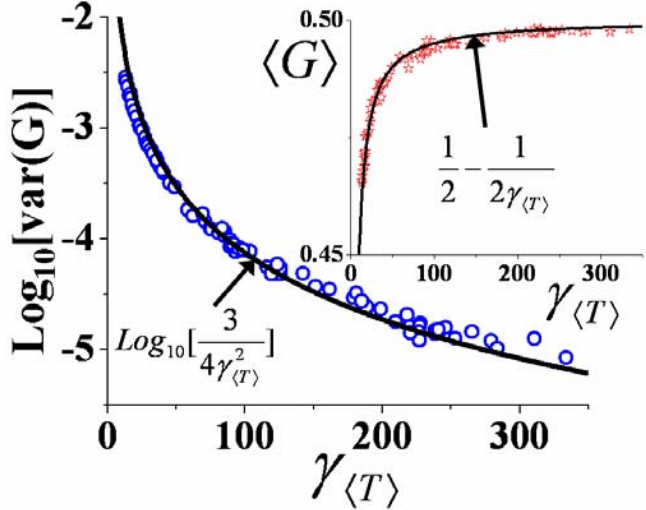

Fig.5: The evolution of the variance of the experimentally determined $P(G;\gamma)$ distributions i.e. $\text{var}(G)$ (shown as the blue circles) for increasing values of $\gamma_{\langle T \rangle}$ is plotted on a logarithmic scale. **Inset:** The evolution of the mean of the experimentally determined $P(G;\gamma)$ distributions i.e. $\langle G \rangle$ (shown as the red stars) for increasing values of $\gamma_{\langle T \rangle}$ is shown. The solid black lines represent the leading terms in Eq.(3) and Eq.(4), and constitute zero-parameter fits.

In Fig. 5, we again employ the sliding frequency window of width 1 GHz to test the asymptotic ($\gamma \gg 1$) relations for the mean $<G>$ (Eq. 3) and variance var($G$) (Eq. 4) of $P(G;\gamma)$ as a function of dephasing (loss) parameter $\gamma$. As before, we determine the value of $\gamma_{\langle T \rangle}$ for each frequency window data set that runs from 3 GHz to 18 GHz for the three Loss Cases-0, 1 and 2. We then determine the corresponding values of the mean and variance of the corresponding conductance distributions $P(G;\gamma)$ of each frequency window. In the inset of Fig. 5, each red star indicates the experimentally estimated mean value of G (i.e., $\langle G \rangle$) for the corresponding value of $\gamma_{\langle T \rangle}$. The standard deviation about the experimentally determined mean is of order $10^{-5}$. We observe that as $\gamma_{\langle T \rangle}$ increases, the red stars asymptotically approach the classical value of $\langle G \rangle = 1/2$. The solid black curve represents the leading terms in Eq. (3).

The blue circles in Fig. 5 show a similar analysis for the variance (var($G$)) of the normalized conductance distributions $P(G;\gamma)$ as a function of $\gamma$. The solid black curve represents the leading term in Eq. (4). We observe that the blue circles closely follow the functional approximation for the theoretical curve (Eq. (4)) for the range of $\gamma_{\langle T \rangle}$ values from about 18 to about 330, with no adjustable parameters.

## VI. Summary and Conclusions

The results discussed in this paper provide experimental evidence in support of the theoretical arguments proposed by [15] and the hypothesis that Random Matrix Theory provides a good description of the conductance fluctuation statistics in a ballistic chaotic quantum-dot in the presence of dephasing. We have shown that in the "locally weak absorbing limit" as discussed in [15], the dephasing parameter can be related to the cavity loss parameter. We have derived an empirical linear relation between $\gamma$ and the cavity loss-parameter $k^2/(\Delta k_n^2 Q)$ based on our experimental data. The finite conductivity of the metallic walls of the cavity translates to a minimum-possible experimentally accessible $\gamma$-value of about 11 for our experiments (at least for the present cavity geometry and temperatures of $-78.5°C$ and above). We have shown that our experimentally determined conductance distributions and the asymptotic analytic functional forms for the PDF of G ($P(G)$), its mean value ($\langle G \rangle$) and variance (var($G$)) are in good agreement over a broad range of large $\gamma$ values. This also establishes the microwave analog as a method to study detailed theories of non-interacting quantum transport and de-coherence in quantum coherent systems.

We acknowledge useful discussions with R. Prange and S. Fishman, as well as comments from P. Brouwer, P. Pereyra and T. Seligman. This work is supported by the DoD MURI for the study of microwave effects under AFOSR Grant

F496200110374, AFOSR DURIP Grants FA95500410295 and FA95500510240, and by the Israel/U.S.A. Binational Science Foundation.